\documentclass[aps,twocolumn,prl,showpacs,superscriptaddress]{revtex4-1}

\usepackage{color}
\usepackage[english]{babel}
\usepackage{graphicx}
\usepackage{epstopdf}
\usepackage{amssymb}
\usepackage{amsmath}
\DeclareMathOperator{\arccot}{arccot}
\usepackage{bm}

\begin{document}
\title{Nonlinear photoionization of transparent solids: \\ a nonperturbative theory obeying selection rules}			% used by \maketitle
\author{N. S. Shcheblanov}
\email{nikita.shcheblanov@polytechnique.edu}
\affiliation{Centre de Physique Th{\'e}orique, CNRS, {\'E}cole Polytechnique, F-91128 Palaiseau, France}
\affiliation{Laboratoire des Solides Irradi\'es CEA-CNRS, Ecole Polytechnique, F-91128 Palaiseau, France}
\author{M. E. Povarnitsyn}
\affiliation{Joint Institute for High Temperatures, RAS, 13 Bld.~2 Izhorskaya str., 125412 Moscow, Russia}
\affiliation{Centre de Physique Th{\'e}orique, CNRS, {\'E}cole Polytechnique, F-91128 Palaiseau, France}
\author{P. N. Terekhin}
\affiliation{National Research Centre `Kurchatov Institute', Kurchatov Sq. 1, 123182 Moscow, Russia}
\author{S. Guizard}
\affiliation{Laboratoire des Solides Irradi\'es CEA-CNRS, Ecole Polytechnique, F-91128 Palaiseau, France}
\author{A. Couairon}
\affiliation{Centre de Physique Th{\'e}orique, CNRS, {\'E}cole Polytechnique, F-91128 Palaiseau, France}

\date{\today}

\begin{abstract}
We provide a nonperturbative theory for photoionization of transparent solids. 
By applying a particular steepest-descent method, we derive analytical expressions for the photoionization rate within the two-band structure model, which consistently account for the {\it{selection rules}} related to the parity of the number of absorbed photons ($odd$ or $even$). We demonstrate the crucial role of the interference of the transition amplitudes (saddle-points), which in the semi-classical limit, can be interpreted in terms of interfering quantum trajectories. Keldysh's foundational work of laser physics [Sov. Phys. JETP 20, 1307 (1965)] disregarded this interference, resulting in the violation of {\it{selection rules}}. We provide an improved Keldysh photoionization theory and show its excellent agreement with measurements for the frequency dependence of the two-photon absorption and nonlinear refractive index coefficients in dielectrics.
%
%Multiphoton experiments on transparent solids and the corresponding perturbation theory highlighted spectral features, originating in {\it{selection rules}}, related to the parity of the number of absorbed photons ($odd$ or $even$). These features are not reproduced by the Keldysh theory foundational work of laser physics [Sov. Phys. JETP 20, 1307 (1965)]. In order to understand the origin of this discrepancy, we revise the approximations of the Keldysh theory. We find that one of Keldysh's assumptions within summation of saddle-points is responsible for a inconsistency in the final expression for the photoionization rate. We present a solution by appropriately accounting for the {\it{selection rules}} in Keldysh's theory. However, in the semi-classical limit, this summation can be interpreted in terms of interfering quantum trajectories. Thus, the disregard of the role of phase due to interference violates the {\it{selection rules}}. It is shown that a correction to the Keldysh theory related to steepest-descent method reduces, finally, the results to the equivalent results of the perturbation theory and yields excellent agreement with experimental measurements.
\end{abstract}

\maketitle

%\section{Introduction}
%
The permanent development of high-power pulsed lasers continues to attract attention to multiphoton processes, predicted by Dirac~\cite{dirac27} and
G{\"o}ppert-Mayer~\cite{gp31}. Theses processes are important for a number of applications like spectroscopy~\cite{axt04,bovensiepen16}, photoemission studies~\cite{damascelli03,pazourek15,bovensiepen16}, high harmonic generation in solids~\cite{td17,tamaya16,vampa15,vampa14,higuchi14}, or optical communications~\cite{freymann10}. In particular, the spatially confined excitation produced by two-photon absorption (2PA) is useful for three-dimensional data storage and imaging~\cite{parthenopoulos89,denk90,cumpston99}. Recently, a possible way towards two-photon semiconductor lasers has been proposed~\cite{reichert16}. These successes have roused the interest in exploring applications based on three-photon absorption (3PA)~\cite{he02} and higher order multiphoton processes~\cite{kazansky14,kerse16}.

In 1964, Leonid Keldysh developed a cornerstone theory~\cite{keldysh65} dedicated to multiphoton processes. While experimental data for the multiphoton absorption coefficient were favorably compared to Keldysh's formula for the ionization probability (see Eq.~(37) in Ref.~\cite{keldysh65}), several authors point out a discrepancy by as much as an order of 
%magnitude, {\it if not a } lack of spectrally resolved measurements~\cite{liu78,nathan85,desalvo96}.
magnitude, if not a lack of spectrally resolved measurements~\cite{liu78,nathan85,desalvo96}.
%
%{Measurements of absorption coefficients in solids have been regularly confronted to Keldysh's theory~\cite{liu78,nathan85,desalvo96}. The Keldysh formula (KLD) for the ionization probability (see Eq.~(37) in Ref.~\cite{keldysh65}) compares favorably to measurements, however, it underestimates and overestimates in many cases the experimental data for the multiphoton absorption coefficient by as much as an order of magnitude~\cite{vaidyanathan80,nathan85,desalvo96}. }
%
%Moreover, the experiments on transparent solids and the corresponding perturbation theory solutions~\cite{braunstein64,weiler73} yield ``selection rules" features of the frequency dependency, related to the parity of absorbed photons number ($odd$ or $even$) at multiphoton regime, that are not reproduced by the Keldysh theory~\cite{weiler73,vaidyanathan80,brandi84}.
%
Moreover, experiments were conducted in the class of transparent solids with inversion symmetry allowing for one-photon transition~\cite{elliott57}, and confirmed the frequency dependence predicted by the perturbation theory~\cite{braunstein64,vaidyanathan80,nathan85} for the $l$-photon transition rate as
\begin{equation}
w_l\sim\left\{
  \begin{array}{ll}
  \left (l\hbar\omega- \epsilon_g \right)^{1/2}, & l-\textrm{odd} \\
 \left (l\hbar\omega-\epsilon_g \right)^{3/2}, & l-\textrm{even}
  \end{array}
  \right.
  \label{SR}
\end{equation}
where $\epsilon_g$ is the band-gap. In contrast, the Keldysh theory reduces to expression $w_l \sim \sqrt{l\hbar\omega-\epsilon_g}$ for $l$-odd and $l$-even \cite{keldysh65}, therefore 
%it does not follow Eq.~(\ref{SR})~\cite{vaidyanathan80,nathan85} and 
violating the {\it{selection rules}}~\cite{vaidyanathan80,nathan85}.
Possible reasons for this discrepancy were proposed by Vaidyanathan {\it et al.}~\cite{vaidyanathan80}, who highlighted simplifying assumptions in Keldysh's derivation with regard to the electronic band structures and oscillator strengths. In order to achieve better agreement between theory and measurements, they suggested to replace the {\it approximate saddle-point integration} in the Keldysh derivation by an exact integration.

%{Possible reasons for this discrepancy were proposed~\cite{vaidyanathan80}: ``This is probably due to the simplifying assumptions made with regard to the electronic energy-band structures and oscillator strengths. Closer agreement between theory and experiment could probably be achieved by replacing the approximate saddle-point integration in the Keldysh derivation by an exact integration.''}

%there have been proposals to obtain an exact Keldysh theory by
%
In this Letter, we revisit the Keldysh theory (KLD). We show that an appropriate modification of one of Keldysh's approximations ensures that the theory, indeed, obeys the {\it{selection rules}}, as perturbation theory does. We perform a detailed comparison of the corrected Keldysh model (cKLD) with recent data on two-photon absorption, yielding an excellent agreement.
%{sKLD}

%\section{Revision}
%
In Ref.~\cite{keldysh65}, the description of the non-perturbative method to derive the expression for the photoionization rate using the Houston wave-functions~\cite{houston40} has been discussed in detail while features such as {\it{selection rules}} at low intensity~\cite{braunstein64}, modulation of photoionization rates with intensity caused by the dynamic Stark effect, and the calculation procedure of matrix elements have not been discussed in full detail. In order to examine the Keldysh approximations, one should draw attention to: a) matrix element approximation and b) details concerning integral calculation.
%{To evaluate the matrix elements, one may use a method similar to the conventional {\it{saddle-point}} method to obtain approximate closed analytical formulas.}
In this connection we refer the reader to the recent Letters~\cite{mcdonald17,zhokhov14} also dedicated to the approximations in Keldysh's theory. These papers deal mostly with approximations of the band structure to unravel the difference between semiconductors and dielectrics. Here, we focus on obeying of the {\it{selection rules}}.

\begin{figure}[ht]
\includegraphics[width=8.2cm]{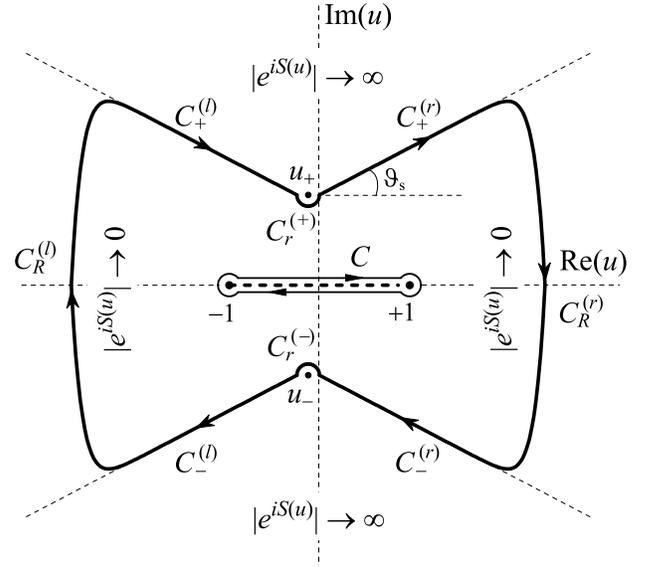}
\caption{The integration contour for $\mathcal{L}_{cv}$ in the $u$ domain is depicted. The poles of $\mathcal{V}_{cv}$ lie at $u=u_{\pm}$ and there is a branch cut between $-1 < u < +1$ on the real axis. The contours $\mathcal{C}_{R}^{(r)}$ and $\mathcal{C}_{R}^{(l)}$ are used to connect the contours $\mathcal{C}^{(r)}_{\pm}$ and $\mathcal{C}^{(l)}_{\pm}$ at infinity.}
\label{Contour}
\end{figure}

%{The saddle-point method can be applied in both the length and velocity gauge. Despite its wide-spread and long-term use in gas (atomic) case --- already Keldysh applied the saddle-point method in the initial work concerned with the limit of small momenta of the outgoing electron and Ref.~\cite{keldysh65} reviews other limiting formulas --- we are not aware of a similar study of the limitation of the saddle-point method in the length gauge in solid-state case.}
%
In order to proceed with the analysis, we quote Eq.~(27) in Keldysh's work~\cite{keldysh65} for the transition rate $w_{pi}$ from an initial state (valence band) to a final state (conduction band) due to the harmonic field $\mathcal{E}_{L}(t) = \mathcal{E}\mathrm{cos}(\omega t)$ with amplitude $\mathcal{E}$ and frequency $\omega$:
\begin{equation}
w_{pi} = \frac{2\pi}{\hbar} \int { \frac{\mathrm{d}\mathbf{p}}{(2\pi\hbar)^3} \left | \mathcal{L}_{cv} (\mathbf{p}) \right |^2 \sum_{l} {\delta(\overline{\epsilon_{cv}(\mathbf{p})}-l\hbar\omega)}},
\label{wpi}
\end{equation}
where the matrix element $\mathcal{L}_{cv}$ can be defined as an integral over a closed contour $\mathcal{C}$ (enclosing the interval $(-1,1)$, see Fig.~\ref{Contour}) in the variable $u = \mathrm{sin}\,\omega t$ (see Eq.~(29), Ref.~\cite{keldysh65}):
\begin{equation}
\mathcal{L}_{cv} (\mathbf{p}) = \frac{1}{2\pi} \oint_{\mathcal{C}} {\mathcal{V}_{cv}(u)e^{iS(u)} \mathrm{d}u},
\label{LCV}
\end{equation}
and $S(u)$ is the classical action:
\begin{equation}
S(u) = \frac {1}{\hbar\omega} \int_{0}^{u} { \frac{\epsilon_{cv}(\mathbf{p}(v))}{\sqrt{1-v^2}} \mathrm{d}v}.
\label{SU}
\end{equation}
The presence of a large factor in the exponent in Eq.~(\ref{LCV}) allows us to calculate the integral $\mathcal{L}_{cv}$ over $u$ by a method similar to the conventional {\it saddle-point} method. Here, we unravel the key aspects of the method allowing us to calculate the PI rate consistently with the {\it{selection rules}}.
%
%{specify the method for calculating the photoionization rate omitted in the original paper~\cite{keldysh65} to unravel the {\it{selection rules}} inconsistency.}

The saddle-points $u_s$ are determined by the condition $\epsilon_{cv}(\mathbf{p}(u))=0$, where the index $s: = \{\pm\}$ specifies one of the special points. However, unlike the conventional {\it{saddle-point}} method, the function $\epsilon_{cv}(\mathbf{p}(u))$ is not analytic and the pre-exponential factor $\mathcal{V}_{cv}(u)$ has poles at these points. The character of the singularities was considered in detail by Keldysh~\cite{keldysh57} and Krieger~\cite{krieger67}. Taking these features into account, we deform the integration contour $\mathcal{C}$ with respect to $u$ as follows (see Fig.~\ref{Contour}).
We deform it from the real axis to the lower and upper half-planes so that it passes around the points $u_s$ along semicircles of infinitesimal small radius $r$ (via the integration paths $\mathcal{C}_r^{(s)}$), goes along the rays $\mathcal{C}_{s}^{(r)}$ and $\mathcal{C}_{s}^{(l)}$, where the contours $\mathcal{C}_{R}^{(r)}$ and $\mathcal{C}_{R}^{(l)}$ are used to connect the contours $\mathcal{C}^{(r)}_{s}$ and $\mathcal{C}^{(l)}_{s}$ at infinity. A simple analysis shows that the integrals $\mathcal{L}_{cv}$ along the $\mathcal{C}_{R}^{(r)}$ and $\mathcal{C}_{R}^{(l)}$ vanishes, and the integrals along the rays $\mathcal{C}_{s}^{(r)}$ and $\mathcal{C}_{s}^{(l)}$ cancel each other (for details see Supplemental Material~\cite{suppl}). In fact, the integration reduces to bypassing singularities with a infinitesimal small radius.

In order to evaluate the remaining integrals, we use $u = u_s + \xi$, and represent the function $S(u)$ in the form:
\begin{equation}
S(u) = \int_{0}^{u_s}+\int_{u_s}^{u_s+\xi}= S(u_s) + S_s(\xi).
  \label{SU1}
\end{equation}
By expanding the pre-exponential factor in Eq.~(\ref{LCV}) near $u_s$, we obtain in the frame of a two-band model~\cite{keldysh57,krieger67} for solids where one-photon transition is allowed~\cite{elliott57}:
\begin{equation}
\mathcal{V}_{cv}(u) \sim \frac{\textrm{sign}(u_s)}{4i(u-u_s)}.
\label{SU2}
\end{equation}
Thus, accounting for $\xi = u - u_s$, we obtain
\begin{equation}
\mathcal{L}_{cv} (\mathbf{p}) \sim \frac{1}{4}\sum_s \textrm{sign}(u_s) e^{iS(u_{s})} \int_{\mathcal{C}_r^{(s)}} {\frac{e^{iS_s(\xi)}}{\xi}\mathrm{d}\xi}.
\label{LCV1}
\end{equation}
In order to complete the integration in Eq.~(\ref{LCV1}), the dispersion law $\epsilon_{cv}(\mathbf{p})$ must be specified. The essential difference from the Keldysh description is, however, the fact that the dispersion law must be specified at this stage rather than at stage of integration over the momentum $\mathbf{p}$ in Eq.~(\ref{wpi}). This is due to the necessity to determine the Stokes (steepest-descent) line angles. Using a certain dispersion law %{$\epsilon_{cv}(\mathbf{p})$}
in $S_s(\xi)$ at $\mathbf{p}=0$, neglecting terms of higher order in $\xi$, and putting $ \xi = r \exp (i\vartheta)$, we find that the steepest-descent lines are the rays corresponding to $\vartheta_s$ and $-\pi-\vartheta_s$. Hence, the angle between the rays determines the final contribution of each saddle-point to~$\mathcal{L}_{cv}$ as follows (for details see Supplemental Material~\cite{suppl}):
\begin{equation}
\mathcal{L}_{cv} (\mathbf{p}) \sim \frac{1}{4}\sum_s \textrm{sign}(u_s)(\pi+2\vartheta_s) e^{iS(u_{s})}.
\label{LCV2}
\end{equation}
The functions $iS(u_{s})$ in the arguments of the exponential functions in Eq.~(\ref{LCV2}) and also the quasienergy $\overline{\epsilon_{cv}}$ in Eq.~(\ref{wpi}) can be calculated exactly (see Supplemental Material~\cite{suppl}). In result, substituting the obtained expression for $\mathcal{L}_{cv}$ in Eq.~(\ref{wpi}) and summing over the momentum, we obtain the final result for the total probability of an interband transition per unit time and per unit volume, see Eqs.~(\ref{wpiK}-\ref{aK}) and (\ref{wpiP}-\ref{aP}).

Keldysh supposed~\cite{keldysh65} that: {\it ``the term in Eq.~(36) } [$iS(u_s)$ in our notations]{\it, which is linear in $x$} [dimensionless momentum]{\it, will henceforth be left out, for when account is taken of both saddle-points it gives rise in $\mathcal{L}_{cv}$ to a rapidly oscillating factor of the type $2 \mathrm{cos}(ax)$, which reduces after squaring and integrating with respect to $x$ to a factor 2, which we can take into account directly in the final answer."} However, we show that this assumption violates the {\it{selection rules}} (see Eq.~(\ref{SR})).
The argument $iS(u_{s})$ of the exponential function in Eq.~(\ref{LCV2}) is calculated allowing for the properties of the functions which determine it in the complex plane. Due to the summation in Eq.~(\ref{LCV2}), the contribution to the integral $\mathcal{L}_{cv}$ from both saddle-points located in the complex plane acquires a phase factor %{which is $-i l\pi$}
(for details see Supplemental Material~\cite{suppl}):
\begin{equation}
\mathcal{L}_{cv} (\mathbf{p}) \sim \left (e^{i\varphi(\mathbf{p})}-e^{-il\pi-i\varphi(\mathbf{p})} \right)\sim\sin\left(\pi l /2+\varphi\right).
\label{LCV3}
\end{equation}
Hence, for solids with an allowed one-photon transition, even-photon absorption is forbidden at $({l\hbar\omega-\epsilon_g}) \, \approx\,0 $ (i.e., $\varphi \,\sim \,0$), as evidenced by the perturbation theory~\cite{braunstein64}. Thus, the correct result is obtained via the proper treatment of the interference of the transition amplitudes and does not require an exact integration.

An interference factor of a similar nature was first obtained by Perelomov {\it{et~al.}}~\cite{perelomov66} in 1966 for the PI rate of {\it atoms}. However, the idea that interfering effect is important for understanding {\it{selection rules}} has been put forward only recently by Popruzhenko {\it{et~al.}}~\cite{popruzhenko02,korneev12} who derived a quantum equation for the PI rate and interpreted it in terms of quantum interference of scattering amplitudes using the self-consistent Born approximation and the Keldysh technique~\cite{popruzhenko02}. Our approach is based on a similar physical picture. Thus, the summation Eq.~(\ref{LCV2}) can also be interpreted in terms of interfering quantum trajectories. A key feature of our approach is simplicity since matrix element can be directly evaluated the two-band model in solids, see Eq.~(\ref{SU2}).
%
%\section{Results and discussion}

In result, we derived a closed-form solution for the photoionization rate in transparent solids within the two-band model obeying the {\it{selection rules}}. The total photoionization rate per unit of volume is given by
\begin{equation}
w_{pi}(\omega)=\sum_{l=l_{pi}}{w^{pi}_l(\omega)}.
\label{sumwpi}
\end{equation}
In the case of the Kane band structure
\begin{equation}
\epsilon_{cv}(\mathbf{p}) = \epsilon_g \left (1 + \frac{\mathbf{p}^2}{m_r\epsilon_g} \right)^{1/2},
\label{kaneBS}
\end{equation}
the corresponding relative PI rate is of the form:
\begin{equation}
w_{l}^{pi}=\frac{4\omega}{9\pi}\left(\frac{m_r\omega}{\hbar\gamma_2}\right)^{3/2}Q^K_l(\gamma,x) \exp \left(-\alpha l_{pi} \right),
\label{wpiK}
\end{equation}
where $\alpha=\pi \left[\mathrm{K}(\gamma_2)-\mathrm{E}(\gamma_2)\right]/\mathrm{E}(\gamma_1)$ and the function
\begin{multline}
Q_l^K(\gamma,x)=\sqrt{\frac{\pi}{2\mathrm{K}(\gamma_1)}}
\phi_l\left(\sqrt{\beta(l-x)}\right) \\ \times
 \exp\left[-\alpha\left(l-l_{pi}\right)\right],
\end{multline}
varies slowly compared with an exponential function. Here, $\gamma=\omega\sqrt{m_r \epsilon_g}/(e\mathcal{E})$ is the Keldysh parameter, $\gamma_1=(1+\gamma^2)^{-1/2}$, $\gamma_2=\gamma(1+\gamma^2)^{-1/2}$, $\beta = \pi^2 / \left [ 2\mathrm{K}(\gamma_1)\mathrm{E}(\gamma_1) \right]$, $\epsilon_g$ is the energy gap, $x=\bar\epsilon_g/\hbar\omega$, $\bar\epsilon_g=2\epsilon_{g} \mathrm{E}(\gamma_1)/(\pi\gamma_2)
$, $l_{pi}=[x+1]$ (where the symbol [$x$] denotes the integer part of a number $x$), the functions $\mathrm{K}$ and $\mathrm{E}$ are the complete elliptic integrals of the first and second kind, and the function $\phi_l$ is
\begin{equation}
\phi_l(z)=e^{-z^2}\int_0^z\sin^2\left(\frac{\pi l}{2}+a y\right)e^{y^2}\mathrm{d}y,
\label{dowson}
\end{equation}
and
\begin{equation}
%a=\sqrt{\frac{\hbar\omega\mathrm{K}(\gamma_1)}{2\pi\epsilon_g\gamma_2}}\left[\arctan\left(\frac{1}{2\gamma}-\frac{\gamma}{2}\right)-\frac{\pi}{2}\right].
a=\sqrt{\frac{\epsilon_g\mathrm{K}(\gamma_1)}{2\pi\hbar\omega\gamma_2}} \arccot \left(\frac{1}{2\gamma} - \frac{\gamma}{2}\right).
\label{aK}
\end{equation}
%
%We also correct a misprint in the expression for the total rate Eq.~(XX)~\cite{keldysh65} which, unfortunately, has been reproduced in many publications.%
%
In the case of a parabolic band structure
\begin{equation}
\epsilon_{cv}(\mathbf{p}) = \epsilon_g \left (1 + \frac{\mathbf{p}^2}{2m_r\epsilon_g} \right),
\label{parabolicBS}
\end{equation}
the total PI rate is also given by Eq.~(\ref{sumwpi}) and the corresponding relative PI rate is of the form:
\begin{equation}
w_{l}^{pi}=\frac{\omega}{4\pi}\left(\frac{m_r\omega}{\hbar}\right)^{3/2}Q_l^P(\gamma,x) \exp \left(-\hat\alpha l_{pi}-\Theta x \right),
\label{wpiP}
\end{equation}
where $\hat\alpha=\left[2\sinh^{-1}(\sqrt{2}\gamma)-\hat\beta\right]$, $\hat\beta=2/\sqrt{1+1/2\gamma^2}$, $\Theta=2\hat\beta\gamma^2/(1+4\gamma^2)$, $\bar\epsilon_g = \epsilon_g \left [1+{1}/({4\gamma^2}) \right]$, and
%$\bar\epsilon_g = \epsilon_g \left(1+\frac{1}{4\gamma^2} \right)$
\begin{equation}
Q_l^P(\gamma,x)=\sqrt{\frac{2}{\hat\beta}} \phi_l\left(\sqrt{\hat\beta(l-x)}\right) \exp\left[-\hat\alpha\left(l-l_{pi}\right)\right]
\end{equation}
and the function $\phi_l$ is the same as in Eq.~(\ref{dowson}) with
\begin{equation}
  a=\sqrt{\frac{2\epsilon_g}{\hbar\omega\hat\beta}}\left(-\frac{1}{\gamma}+\frac{2\sqrt{2}}{\hat\beta}\right).
  \label{aP}
\end{equation}

A simple analysis shows that the rates, Eqs.~(\ref{wpiK}) and (\ref{wpiP}), are reduced to Eq.~(\ref{SR}) via a small {\it z}-expansion of $\phi_l$:
%{in Eq.~(\ref{dowson})}
%
\begin{equation}
\phi_l\sim\left\{
  \begin{array}{ll}
  \left (l\hbar\omega - \bar\epsilon_g \right)^{1/2} & ,~ l-\textrm{odd} \\
 \left(l\hbar\omega - \bar\epsilon_g \right)^{3/2} & ,~ l-\textrm{even}
  \end{array}
  \right.
  \label{SR1}
\end{equation}
i.e., obey the {\it{selection rules}} for any band structure approximation, Kane or parabolic, hence, agree with the perturbation theory at low intensities ($\gamma \gg 1$).
%
%{An easy but important check on any theory of multiphoton absorption is to require that the results agree with the perturbation theory for low light intensities. We present here an expression corresponding to $\gamma \gg 1$ in order to confirm the {\it{selection rules}} : ...}
%
\begin{figure}[ht]
\includegraphics[width=8.5cm]{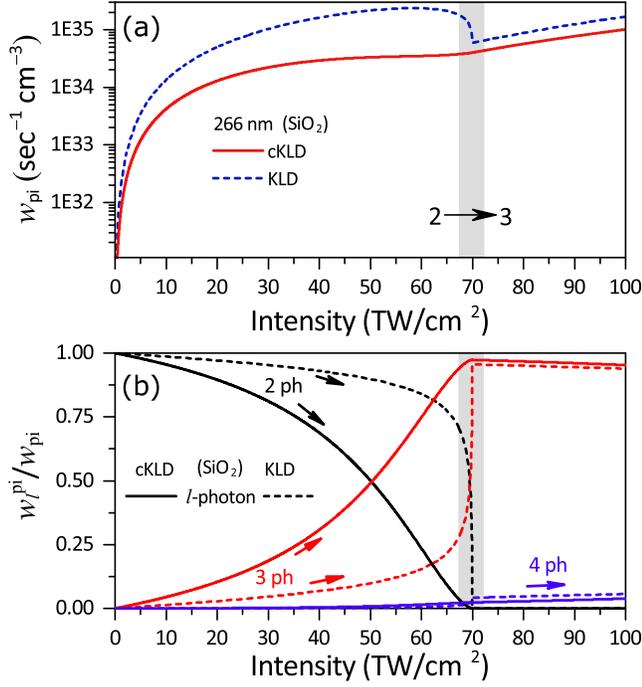} %fig3_new
\caption{Comparison between the original Keldysh and the corrected Keldysh formulas of photoionization rates as functions of laser intensity at 266~nm wavelength in SiO$_2$. Solid curves correspond to cKLD formula and dashed curves reproduce the KLD formula. (a) Total PI rate. (b) Relative contribution of 2- (black), 3- (red) and 4-photon (blue) processes to photoionization. The channel-closure region at $I\approx70$~TW~cm$^{-2}$ is indicated by the grey shaded area, (a) and (b), and horizontal arrow, (a). {\it{Note}}:~\cite{comment}.}
\label{wpikaneBS}
\end{figure}
\begin{figure}[ht]
\includegraphics[width=8.2cm]{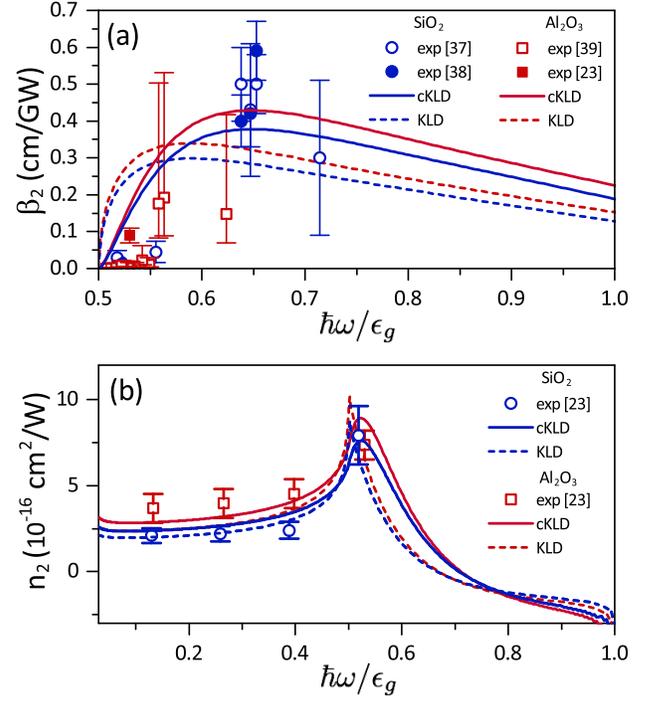}
\caption{Comparison between theory (curves) and experiment (markers) in SiO$_2$ (blue) and Al$_2$O$_3$ (purple). Solid and dashed curves are numerical results via cKLD and KLD theory, respectively.
(a) 2PA coefficient $\beta_2$. Fused silica (empty circles) and crystalline quartz (filled circles)~\cite{nikogosyan02,nikogosyan03}, Al$_2$O$_3$~\cite{leedle17} (empty squares) and~\cite{desalvo96} (filled square).
(b) Experimental and numerical nonlinear refractivity index $n_2$. Experimental data are taken from~\cite{desalvo96}: empty circles for SiO$_2$ and empty squares for Al$_2$O$_3$.
{\it{Notes}}:~\cite{comment,remark}.
%{Shaded areas correspond to frequencies beyond the validity range of the Sellmeier dispersion relation for $n_0(\omega)$.}
}
\label{b2n2}
\end{figure}

In order to compare the results of the corrected theory with experiment, we choose SiO$_2$ and Al$_2$O$_3$, two highly relevant materials for industrial applications. The band structure of these wide-band gap insulators~\cite{waroquiers13,french90} can be well approximated by a two-band model with only two parameters, the reduced mass $m_r$ and the band gap $\epsilon_g$. We use $m_r=0.9~m_0$ and $\epsilon_g=9.0$~eV for SiO$_2$, and $m_r=0.35~m_0$ and $\epsilon_g=8.8$~eV for Al$_2$O$_3$.
%We use $m_r=0.6~m_0$ and $\epsilon_g=9.0$~eV for SiO$_2$~\cite{waroquiers13}, and $m_r=0.2~m_0$~\cite{shu92} and $\epsilon_g=8.8$~eV for Al$_2$O$_3$~\cite{french90}.

Fig.~\ref{wpi}(a) shows the dependence of the PI rate per unit volume on laser intensity, for SiO$_2$ at the laser wavelength 266~nm, as calculated from the KLD theory and the cKLD model.
%It is well known that the Keldysh formula presents cusps associated with a fast decrease of the photoionisation rate over a narrow intensity range around transitions.
The cusp at $\approx70$~TW/cm$^2$ is the signature of 2PA to 3PA transition in the Keldysh formula, and is due to the energy shortage as the electron ponderomotive energy grows up with increasing laser field amplitude (the dynamic Stark effect), and thus, the probability of photoionization decreases sharply highlighting the signature of channel-closing~\cite{paulus01,zimmermann17}. As can be seen in Fig.~\ref{wpi}(a), this cusp is no longer present in our corrected cKLD model reflecting the proper {\it{superposition}} of channels, 2PA and 3PA, respectively.
%{\it{selection rule}} requirements. indicating that it originates from a mathematical artefact due to assumptions in the KLD theory.
 In Fig.~\ref{wpi}(b) we evaluate the relative contribution of multi-photon processes (channels) to the total photoionization rate. For the KLD model, the contribution of 2PA vanishes at $\approx70$~TW/cm$^2$, i.e. the channel closes, while the contribution of 3PA abruptly increases. For the cKLD model, a smooth transition from 2PA to 3PA is obtained: the contribution of 3PA compensates for the attenuation of the 2PA process. % in the vicinity of a channel-closing.

By taking into consideration only the 2PA process, that is valid in the limit of low laser intensities, we compare theoretical predictions for the 2PA coefficient $\beta_2$ calculated from the KLD and cKLD models with measurements for SiO$_2$~\cite{nikogosyan02,nikogosyan03} and recent data for Al$_2$O$_3$~\cite{leedle17}, see Fig.~\ref{b2n2}(a). The improved Keldysh model, cKLD matches better with the experimental findings, especially, in the vicinity of the transition from 2PA to 3PA ($\hbar\omega/\epsilon_g\approx0.5$), where the Keldysh model overestimates the absorption rate by a factor of $\sim10$, and further highlights and confirms the {\it{selection rules}} signature.
The application of the Kramers-Kronig relation to the imaginary part of the permittivity
%to the real part of the conductivity $\sigma^{pi}(\omega)$
gives the frequency dependence of the complex dielectric function $\varepsilon(\omega)=\varepsilon_r(\omega)+i\varepsilon_i(\omega)$, and thus, allowing as to derive the dispersion curves of the nonlinear refractive index $n_2(\omega)$:
%$$n_2(\omega)I= \sqrt{{\varepsilon_r(\omega)+|\varepsilon(\omega)|}} / \sqrt{2}-n_0(\omega),$$
 $$n_2(\omega)I=\operatorname{Re}\left(\sqrt{\varepsilon(\omega)}\right)-n_0(\omega),$$
where $n_0(\omega)$ is the linear index approximated by three-term Sellmeier dispersion equation for SiO$_2$~\cite{malitson65}, Al$_2$O$_3$~\cite{malitson62},
% taken at 266~nm to be 1.59 and 1.83 for SiO$_2$~\cite{malitson65}, Al$_2$O$_3$~\cite{malitson62}, respectively,
and $I=\epsilon_0cn_0\mathcal{E}^2/2$ is the laser intensity in the bulk. Dispersion curves $n_2$ are shown in Fig.~\ref{b2n2}(b), where we present the comparison of the cKLD and KLD models with measurements~\cite{desalvo96}, demonstrating again excellent agreement.
As can be seen, both models give similar behaviour except for the sharp peak (resonance-like behaviour) at half the band gap energy, where Keldysh's model exhibits a cusp originating from the omission discussed above,
%{related to the {\it{interference}} of the transition amplitudes and leading to violation of the {\it{selection rules}} }
whereas the corrected model yields a significant improvement.
%
%\section{Summary}

In this Letter we revise the Keldysh approximations and reveal that after the appropriate correction the Keldysh theory indeed obeys the {\it{selection rules}} and reduces to the equivalent results of perturbation theory. We also demonstrate that the {\it{selection rules}} can be understood as a classical effect caused by interference of quantum trajectories. In order to remedy the Keldysh omission, we propose a simple correction taking into account such interfering effect. The results yield excellent agreement with experimental measurements of the two-photon absorption coefficient $\beta_2$ as well as nonlinear refractive index $n_2$ for materials Al$_2$O$_3$ and SiO$_2$.
%
%In this work it has been demonstrated that the MSD was not correctly employed in
%
%\section{Acknowledgements}

The authors thank S. Popruzhenko and N. Shvetsov-Shilovski for useful discussions. MEP acknowledges support of the CNRS. NSS and MEP are grateful to the Russian Foundation for Basic Research (project No. 16-02-00266) for financial support.
%
%\bibliography{biblio}

\begin{thebibliography}{46}
\bibitem [{com()}]{comment}%
  \BibitemOpen
  \href@noop {} {}\bibinfo {note} {The electron spin is taken into
  account.}\BibitemShut {Stop}%
\end{thebibliography}


\end{document}